# A Dual-SOM Learner Model of Instructional Efficiency within Cognitive Load Theory†

**Spyridon Revithis**

School of Computer Science and Engineering
University of New South Wales, NSW Sydney, Australia
revithiss@cse.unsw.edu.au

## Abstract

This paper presents a novel methodological framework, integrating a dual self-organising map (SOM) neural network architecture with Cognitive Load Theory (CLT), for assessing instructional efficiency within learning environments. By representing curriculum content as a directed acyclic graph of topic dependencies, the proposed model captures collective learning behaviour across two core dimensions: academic performance and subjective/introspective mental effort. During SOM training, representative data sets of successful learner trajectories are compressed into two topologically ordered, identically structured neural networks: PSOM (performance space) and ESOM (mental effort space). By pairing these behavioural maps, the model serves as a decision-support tool that enables instructors to simultaneously pinpoint structural instructional design defects, such as hidden topic prerequisites, and evaluate collective cognitive load profiles. The implementation and performance of the PSOM component (and, by extension, its symmetrical twin, ESOM) is linked to an earlier modelling study employing a SOM with identical structure and encoding semantics situated in the same domain (i.e., a learning environment). This paper combines the study of behavioural phenomena with theoretical work on human cognitive architecture via SOM neurocomputational cognitive behavioural modelling.

**Keywords**: Self-Organising Maps; Neural Network Modelling; Cognitive Load; Working Memory; Instructional Design.

**Acknowledgments**: Extending credit and appreciation to W. H. Wilson and N. Marcus, who contributed to the foundational publication of this work (Revithis, Wilson, and Marcus, 2010).

† This paper is a substantially revised and expanded version of (Revithis, Wilson, and Marcus, 2010).

## 1 Introduction

Studying how people learn and acquire knowledge has long been a focal point in education, driven by an enduring demand for the efficient dissemination and assimilation of knowledge (Sweller, 1999; Sweller, van Merriënboer, & Paas, 2019). Despite the pervasive application of ever-expanding digital technologies and continuously developing educational methods, bridging the distance between learning behaviour and instructional design remains a constant challenge. There are numerous studies on intelligent adaptive computational systems in education based on AI methodologies, web engineering, and cloud-based platforms (e.g., Wenger, 1987; Brusilovsky, 1999; Shang, Shi, and Chen, 2001; Ben-Naim, Marcus, and Bain, 2007; Kabudi, Pappas, and Olsen, 2021; du Boulay, Mitrovic, and Yacef, 2023), demonstrating the ubiquitous need for measuring instructional efficiency.

The work presented here is an extension of early research on behavioural modelling of human learners (Revithis, 2001) and its e-learning applicability prospects (Shi, Revithis, and Chen, 2002). It introduces a novel approach to assessing instructional design, based on collective (rather than individual) working memory load, within the framework of Cognitive Load Theory (CLT) (Sweller, 1988; Sweller, 1994; Sweller, van Merriënboer, and Paas, 1998; Sweller, Ayres, and Kalyuga, 2011; Sweller, van Merriënboer, and Paas, 2019; Sweller, 2025).

In the next section, the characteristics of learning environments are examined in relation to human cognitive architecture, and the role of working memory load is analysed. In Section 3, a behavioural learner model based on a dual-SOM neural network is introduced and described in terms of its use as a collective cognitive load indicator. Section 4 presents key implementation details as well as evaluation aspects of the dual-SOM component, and the paper concludes by discussing operational limitations and future research directions in Section 5.

## 2 Human Learning

Learning behaviour can be viewed as being constrained and shaped by human cognitive architecture. Because these cognitive structures dictate how information is processed and stored, instructional design largely determines the functional specifications for the acquisition of knowledge within learning environments.

### 2.1 Learning Environments

A generic *learning environment* consists of an *instructor* and a varying number of *learners*, with the central goal of enabling learners to acquire predetermined domain knowledge. Traditionally, this process is facilitated through structured interactions between the learners and the instructor.

By convention, a learning environment corresponds to a unit of study (such as introductory algebra), the learning content of which is delivered across a sequence of instructional sessions. Each session introduces a thematically distinct *topic* (for example, equation preliminaries, mathematical operations and their properties, or solving quadratic equations). The structural organisation of this learning content is formalised as a *dependency graph* of topics, where each topic contains a subset of the domain content and requires zero or more prerequisite topics. This prerequisite structure reflects the prior knowledge necessary to engage with subsequent material; attempting a topic before completing its prerequisites violates the dependency constraints and results in a deficit of the foundational knowledge required to assimilate new information. A learner is deemed to have successfully completed a topic upon achieving adequate performance in the corresponding topic assessment. Figure 1 illustrates an example dependency graph for hypothetical five-topic learning content. In this structure, Topic Z serves as a prerequisite for both Topic S and Topic J, while Topic Q and Topic B act as additional prerequisites for Topic J. Conversely, Topics Z, Q, and B have no prerequisite dependencies and represent valid entry points into the learning environment.

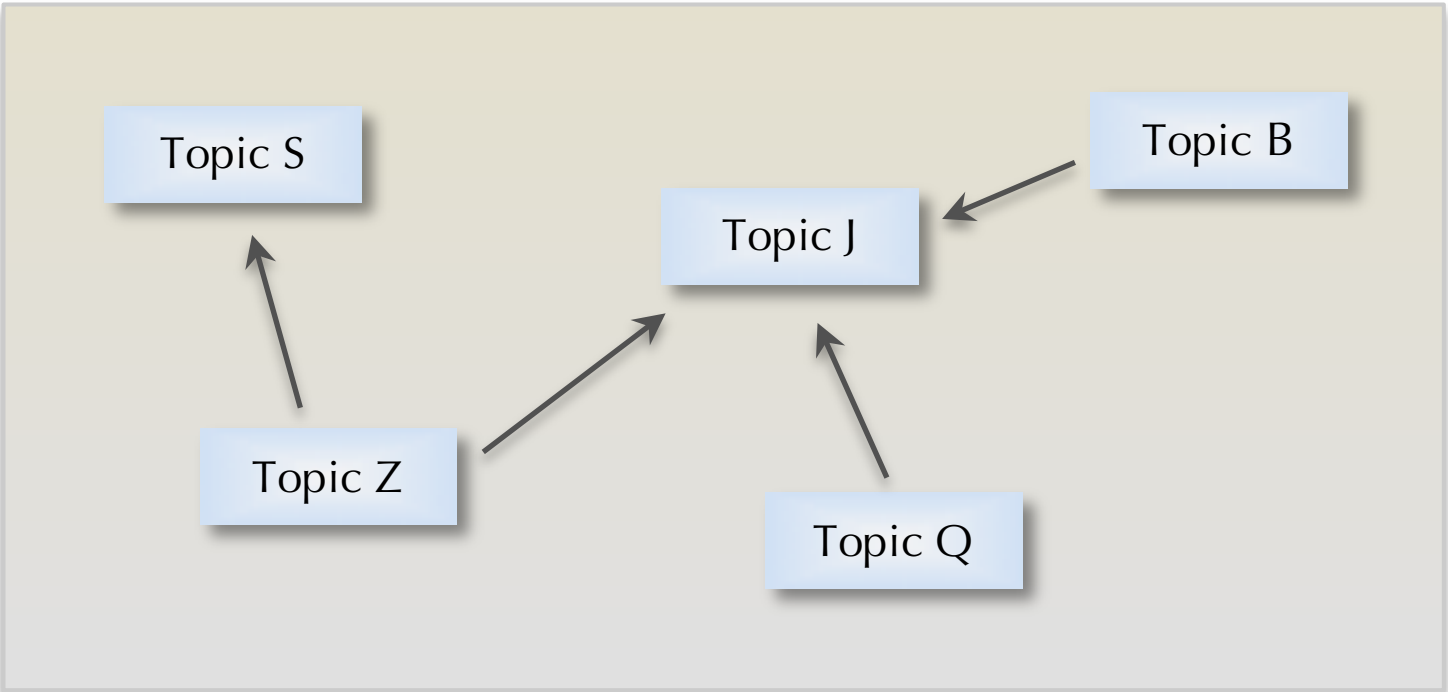

**Figure 1**: Example dependency graph of five-topic learning content within a learning environment.

By design, each topic has a high level of content self-sufficiency, enabling learners to focus their effort on a single topic without engaging in multiple topics simultaneously. Although conceptually self-contained, topics are often not independent; many require prerequisite knowledge distributed across the curriculum. This network of content dependencies, formalised by the dependency graph, dictates a valid ordering within the learning environment and establishes the baseline structure for instructional design.

Learners entering an instructional environment possess varying degrees of prior *background knowledge*. According to *schema theory* (Bartlett, 1932; Minsky, 1975; Bobrow and Winograd, 1977; Larkin et al., 1980; Chi, Glaser, and Rees, 1982; Mayer, 1992) knowledge is organized within long-term memory into categorical, retrieval-oriented units termed schemas, which perform lossy compression on complex information. These schematic structures play a critical role in the acquisition and application of new knowledge (Kalyuga, 2010; Sweller, van Merriënboer, and Paas, 1998), enabling working memory to interpret novel inputs against existing representations.

Whether a learner's prior knowledge is adequate for a given instructional environment is typically assessed through diagnostic screening upon entry. However, diagnostic imperfections may allow individuals with an insufficient knowledge base to proceed undetected. This knowledge mismatch results either in the complete absence of required schemas or in a structural incompatibility between the learner's existing schemas and the new domain content.

Typically, the learners go through timely and systematic evaluation so that the instructor can obtain qualitative information on their acquired knowledge. The overall learner's feedback provides particularly valuable information on the effectiveness and efficiency of the instructional design of the learning environment.

## 2.2 Working Memory and Cognitive Load

The set of all valid sequences for successfully completing the topics of a learning environment $E$, bounded by the constraints of the dependency graph, constitutes the environment space $S(E)$. As the number of topics increases, the cardinality $|S(E)|$ grows factorially; for $n$ unconstrained topics, $|S(E)| = n!$. Graph-theoretically (Cormen et al., 2009), the dependency structure is formalised as a directed acyclic graph (DAG), where each topic corresponds to a node and each prerequisite dependency represents a directed edge. A valid sequence to complete all topics thus corresponds to a topological ordering of the DAG, where a directed edge from node $j$ to node $i$ requires node $j$ to precede node $i$ in the traversal. If the graph is sparse ($m \ll n$, where m is the number of edges), $|S(E)|$ approximates $(n - m)!$. Conversely, in a strictly linear sequence ($A_1 \rightarrow A_2 \rightarrow A_3 \rightarrow A_4 \rightarrow \ldots \rightarrow A_n$), the DAG edges completely constrain the path, yielding $|S(E)| = 1$. The learner must navigate this constrained space to identify an optimal traversal path, one whose topic sequence maximises knowledge acquisition.

Knowledge acquisition can be impaired in any educational setting. This study adopts an empirically supported framework explaining how and why such impairment occurs, grounded in *working memory* capacity and *cognitive load* limitations (Baddeley, 2012; Miyake and Shah, 1999; Kalyuga, 2006; Baddeley and Hitch, 1974), and will be discussed in the remainder of this section. Learners receive information primarily via their audio-visual sensory system into their sensory memory and then process it while it is temporarily retained in working memory; such processing employs schema-encoded knowledge residing permanently in *long-term memory* (Sweller, van Merriënboer, and Paas, 1998). These fundamental elements of human cognitive architecture point to working memory as a focus of conscious mental functioning. In this architecture, the impact of cognitive load, or the associated amount and impact of mental effort on working memory (Sweller, 1994) on knowledge acquisition is substantial, as will be discussed next.

Working memory *capacity* is subject to an intrinsic constraint, permitting only a few chunks of information to be processed simultaneously (Miller, 1956; Cowan, 2001, 2010). In this context, a ‘chunk’ denotes a discrete operational unit handled by working memory as a single element rather than an unorganized set of disparate inputs. This limitation aligns with recent empirical findings from the Time-Based Resource Sharing (TBRS) framework, which show that heavy processing demands consume the attentional resources needed to refresh working memory traces and encode item order into long-term memory (Baggini and Ricciardelli, 2025). From a CLT perspective, when continuous processing monopolises central attention, too few temporal resources remain for germane processing and schema acquisition. In addition to intrinsic capacity constraints, the schemas residing in long-term memory significantly influence working memory processing. Cognitive strain arises from the interplay between finite working memory capacity and a lack of compatible or automated long-term memory schemas (Sweller, van Merriënboer, and Paas, 1998). This excessive burden on working memory, defined as high cognitive load, severely impairs knowledge acquisition and serves as a foundational construct in CLT (Sweller, 1994).

Schema-deficient incoming information disrupts processing between working memory and long-term memory due to the absence of relevant long-term memory schemas required to interpret new knowledge (schema referencing failure). Similarly, incoming information with high structural complexity can cause working memory overload. Information complexity is defined here in terms of element interactivity: the degree of semantic interdependence among constituent information elements (Sweller, 1994). In both instances, the functional consequence manifests as a virtual reduction in effective working memory capacity (Figure 2). This element-interactivity perspective aligns with relational complexity metrics, which provide a rigorous neurocomputational approach to information complexity and working memory processing capacity (Halford, Wilson, and Phillips, 1998, 2010).

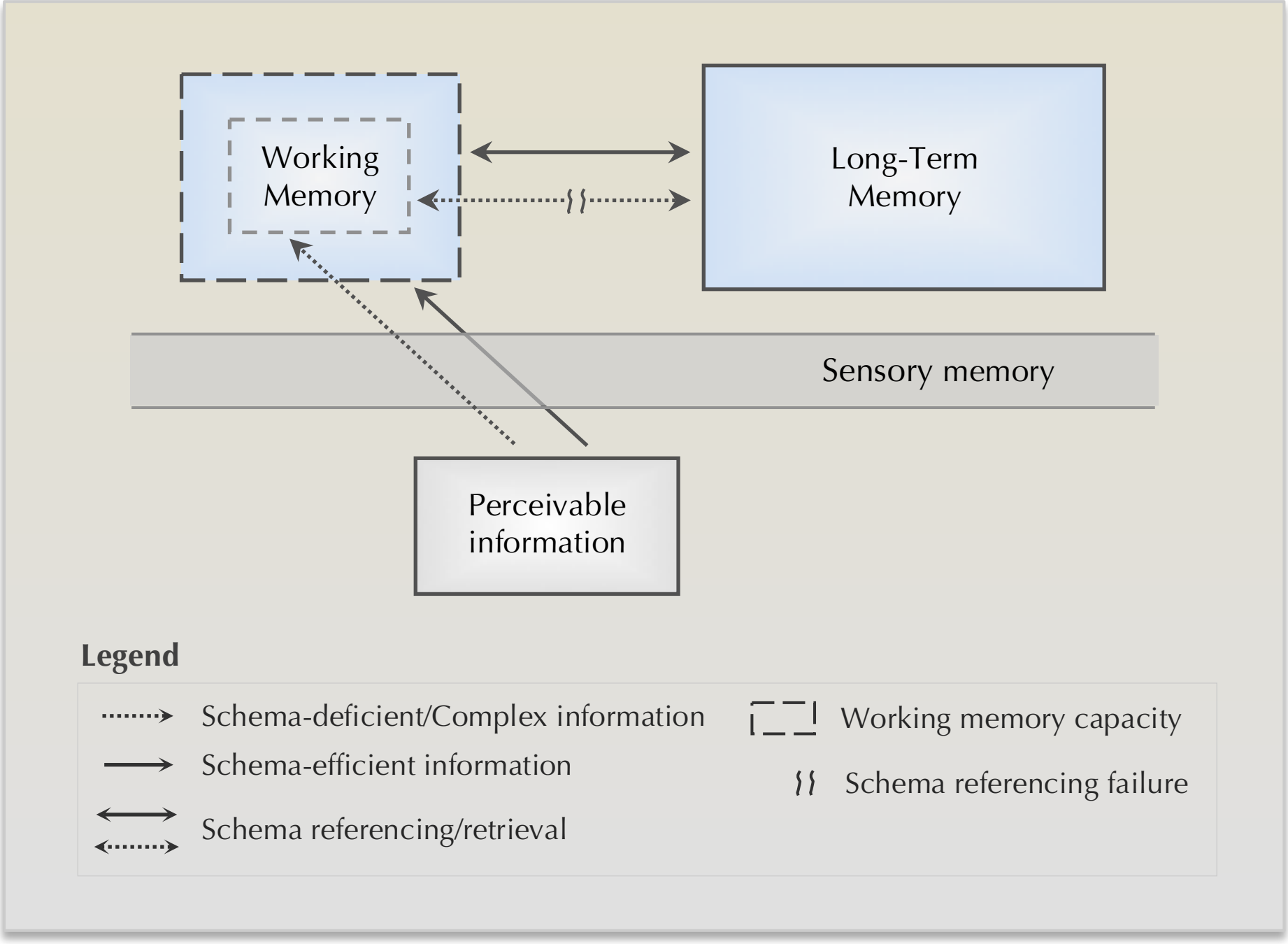

**Figure 2**: Knowledge acquisition as a function of working memory load and schema availability. While absolute physical working memory capacity remains constant, effective processing capacity is virtually reduced when processing schema-deficient or highly complex information.

Within the conceptual framework of CLT, three core dimensions determine the working memory demand during knowledge acquisition: intrinsic load, extraneous load, and germane processing (or germane load). *Intrinsic* cognitive load relates to the inherent structural complexity of the learning content, characterised by element interactivity, relative to the learner's existing schema knowledge base in long-term memory. Suboptimal instructional presentation or redundant task processing generates *extraneous* cognitive load and consumes working memory capacity without facilitating learning.

Historically, CLT framed *germane* cognitive load as an independent, additive third category of load devoted directly to schema construction and automation. In more recent formulations (e.g., Sweller, van Merriënboer, and Paas, 2019; Sweller, 2025), germane load does not impose an additive load in its own right; instead, it functions by redirecting working memory resources away from extraneous activities and toward processing the intrinsic demands of the learning task. Under both perspectives, higher germane processing signifies productive mental effort that fosters schema acquisition, making the distinction between unproductive (extraneous) and productive (germane/intrinsic) effort central to evaluating instructional efficiency.

In Section 3.2, the relationship between working memory capacity and cognitive load is explored in terms of introspective mental effort and measured cognitive performance.

## 3 The Dual-SOM Behavioural Learner Model

Despite the vast variability of human behaviour, domain-specific context and constraints naturally reduce this variability significantly. As will be discussed in Section 3.1, a natural reduction of the behavioural space, which occurs in domain-specific learning, is an initial working assumption in modelling learning behaviour in this study.

In a typical learning environment the instructor reviews the learners' feedback in order to determine the degree of knowledge acquisition and instructional effectiveness. Situated behaviour within a learning environment (that is, domain-specific learning behaviour) can provide the necessary empirical data to construct not only a model of learner performance but also a *behavioural learner model*, allowing for the assessment of instructional design and working memory load as a function of cognitive strain and performance during knowledge acquisition.

The model introduced in this work is based on performance and introspective metrics in order to quantitatively characterise the learners' behaviour and help estimate their cognitive load. The behavioural learner model, which fits the characterisation of a behavioural outcome model (Sun and Ling, 1998), can be implemented as a computational dual *self-organising map* (SOM) neural network, the characteristics and architecture of which will be presented next.

### 3.1 Architecture and Function of the Model

The behavioural learner model, which is described here and linked to CLT in Section 3.2, is based on the novel use of two Kohonen SOM neural networks (Kohonen, 1982, 2001) that map the learners' behaviour within the learning environment.

The first neural network, *PSOM*, represents the learners' cognitive *performance* per topic, whereas the second one, *ESOM*, represents the learners' cognitive *effort* per topic during the same period. Both behavioural maps are formed after PSOM and ESOM undergo training that exposes each neural network to a sizeable data set of representative past cases from learners who have successfully completed all topics. In the case of PSOM, the training set consists of numerical data on *performance in topic* evaluation tests, whereas, in the case of ESOM, it consists of numerically scaled introspective data on the *mental effort* required to successfully complete each topic (Figure 3).

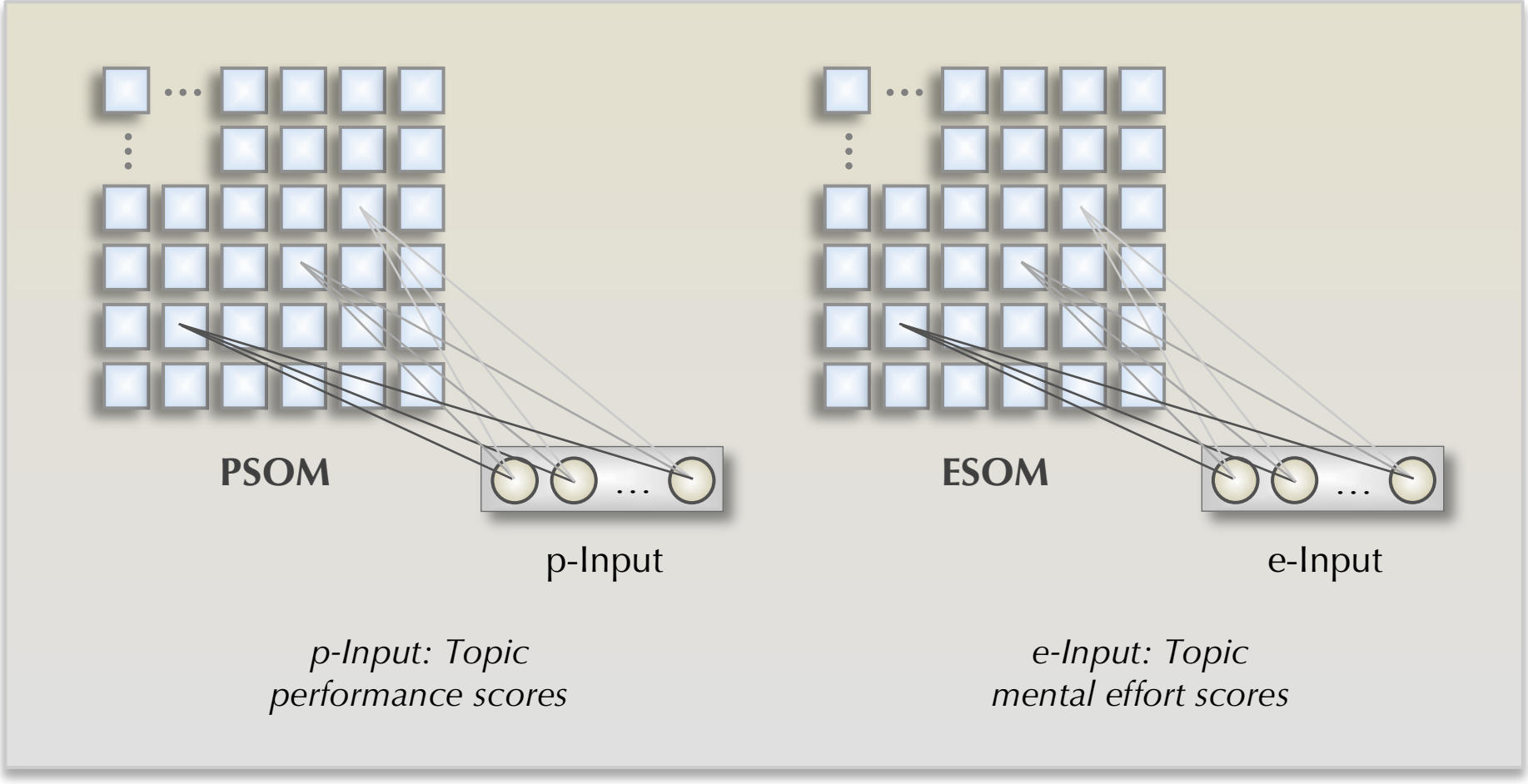

**Figure 3**: Structure and training of PSOM and ESOM networks: e-Input and p-Input connect to all neurons in the map following standard SOM training.

During the training session, each neural network is sequentially presented with performance and mental effort vectors, respectively, in order for the network to converge to a stable state according to the machine learning and termination characteristics of the SOM algorithm; a comprehensive layout of standard SOM analysis is provided by Haykin (2009). The significance and efficiency of the learners' behavioural mapping depends on how statistically representative of the learner population the training data set is.

Assuming that a form of behavioural-space reduction takes place naturally in domain-specific learning, both the performance and mental effort behavioural spaces for any learning environment are generally much smaller than the vast theoretical space of human behaviour, albeit they remain prohibitively large. However, the explosively large space of potential domain-specific learning behaviour can be compressed into a manageable set of behavioural patterns in the form of two topologically ordered behavioural maps, PSOM and ESOM, representing statistically dominant *cognitive behavioural patterns* for performance and mental effort, respectively. Due to the well-established properties of SOM neural networks (Haykin, 2009), it is expected that each map would be a close approximation of the domain behavioural space it compresses, provided that each map has been trained using a statistically representative subset of that behavioural space (Figure 4).

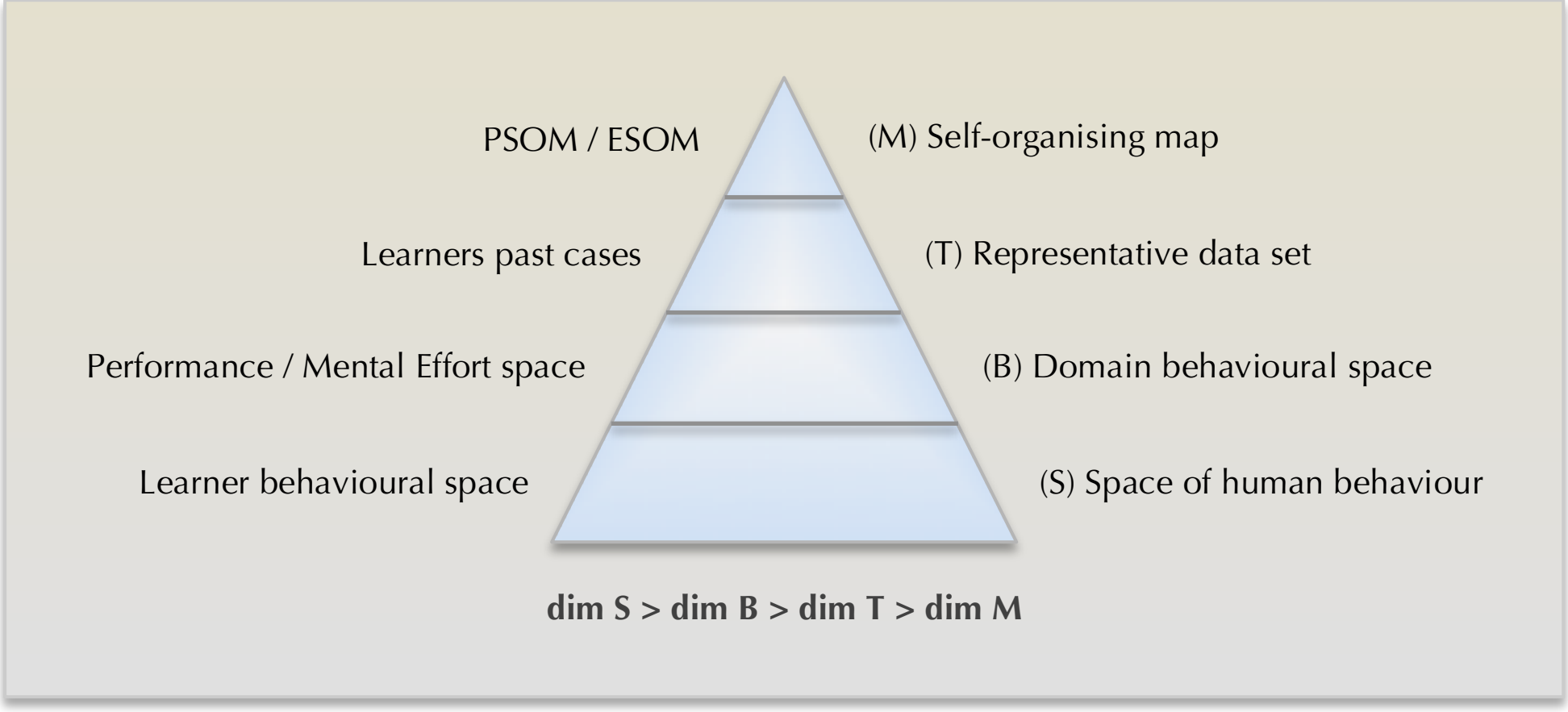

**Figure 4**: PSOM/ESOM Behavioural Space Reduction. Significant space reduction of the original human behavioural space can be achieved via training a SOM network using a statistically representative data set drawn from the domain behavioural space.

Upon completion of training, PSOM constitutes a topological cognitive behavioural map of the learners' performance in each topic of the learning environment. The structure of PSOM is a two-dimensional lattice where each node (neuron) retains a vector pattern of performance. Similarly, the trained ESOM network maps the learners' mental effort in each topic of the environment, with each node in its lattice retaining a vector pattern of mental effort. The size of each lattice is primarily a function of the number of performance (or mental effort) patterns the model designer deems appropriate for the SOM network to represent; for the proposed design, it is in the order of 100 nodes per map. Implementation of the model is discussed in Section 4.

Each node in either map contains not the performance (or mental effort) of a single learner in the environment, but a performance (or mental effort) pattern; that is, it represents the centre of a cluster of similar performance (or mental effort) values for a statistically dominant group of learners. Therefore, each node represents a distinct family of characteristic learning behaviours drawn from the learners' domain behavioural space. Due to the SOM topological property, neighbouring nodes in the map represent similar behaviours.

### 3.2 Cognitive Load Metrics and Instructional Efficiency

The formation of learner behavioural maps enables inferences regarding the learners' cognitive response to the instructional design of the learning environment. Earlier work (Revithis, 2001) has shown that a performance-based SOM learner model can decisively assist in revealing behavioural patterns associated with topic dependencies that are not depicted in the original dependency graph. For instance, learners might tend to perform poorly if they study topics in a certain order, even though that specific ordering does not violate the graph of known dependencies. Hypothetically, in the example in Figure 5, the graph depicts a *hidden dependency*, in which Topic S is also a prerequisite for Topic J; this can be revealed by the trained PSOM behavioural patterns, in which learners tend to perform poorly in Topic J if it is completed before Topic Z and Topic S. Omitting such hidden dependencies from the dependency graph can be attributed to inefficient instructional design, provided that all learners possess sufficient background knowledge before entering the learning environment. Therefore, PSOM can be decisive in determining which ordering of topics is optimal for knowledge acquisition, and which topics present a challenge to the learners.

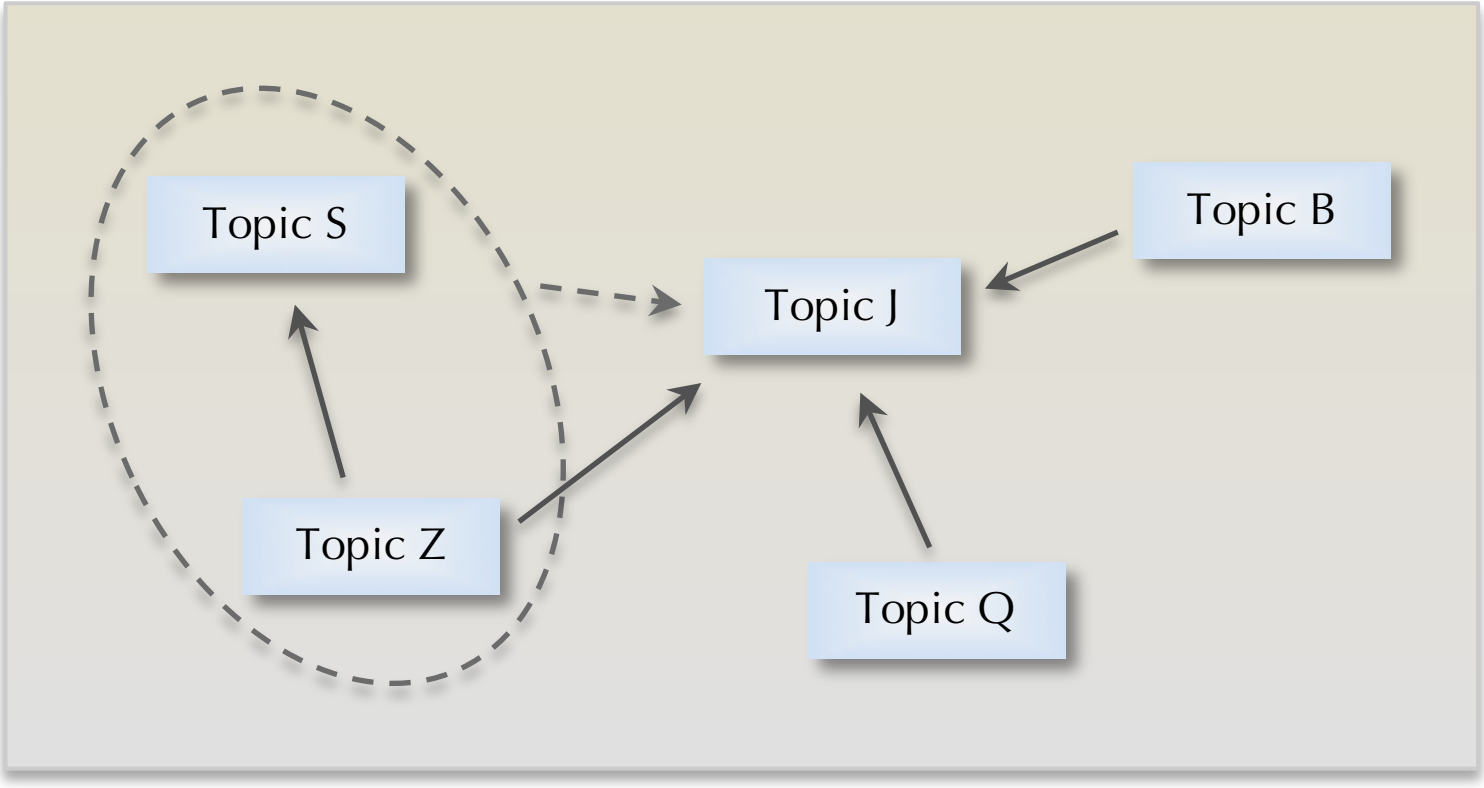

**Figure 5**: Example of a hidden dependency in a dependency graph of five-topic learning content (solid arrows indicate known prerequisites; the dashed oval and arrow denote the hidden dependency between Topic S and Topic J).

Mental workload can be assessed through subjective/introspective, physiological and performance-based methods (Wierwille and Eggemeier, 1993). Among these, it has been shown that the subjective measurement of cognitive load is less intrusive, easier to implement than the physiological measurement approaches, and just as valid (Paas, van Merriënboer, and Adam, 1994; Marcus, Cooper, and Sweller, 1996). Furthermore, it has been suggested that humans can easily introspect on their mental load using a numerical scale (Gopher and Braune, 1984; Leppink et al., 2013), and that the scaling convention does not affect the result of the introspection (Borg, 1978; Hendy, Hamilton, and Landry, 1993). This evidence justifies the inclusion of a subjective/introspective metric component of mental effort (ESOM) in the model.

By combining performance and introspective data in the behavioural model, the learning environment can be evaluated in terms of cognitive load and instructional design. Pairing performance and mental effort scores is consistent with literature recommending such a combined metric for measuring the efficiency of instructional design (Sweller, Merriënboer, and Paas, 1998; Marcus, Cooper, and Sweller, 1996; van Gog and Paas, 2008; van Gog, Hoogerheide, and van Harsel, 2020).

The scores recorded in the trained PSOM and ESOM are critical in determining the learners' cognitive inclinations; based on such data, the instructor can deduce, among other things, where learners usually excel in terms of knowledge acquisition, where they struggle yet still achieve well, where they fail to perform, and where they perform adequately. By combining performance and mental effort norms, the learners' scores can be interpreted through the lens of CLT, yielding meaningful conclusions regarding cognitive load levels in the learning environment under investigation.

In Figure 6, a hypothetical example is presented for a five-topic learning environment. Here, it is assumed that the instructor has accessed the trained PSOM and ESOM components of the model and identified a consistent pattern cluster (i.e., grouping of patterns in each trained SOM). Drawing on the relationship between combined performance-mental effort data and CLT, described earlier in this section and in the preceding Section 2.2, it becomes possible to construct a *cognitive load profile* for each topic in the learning environment.

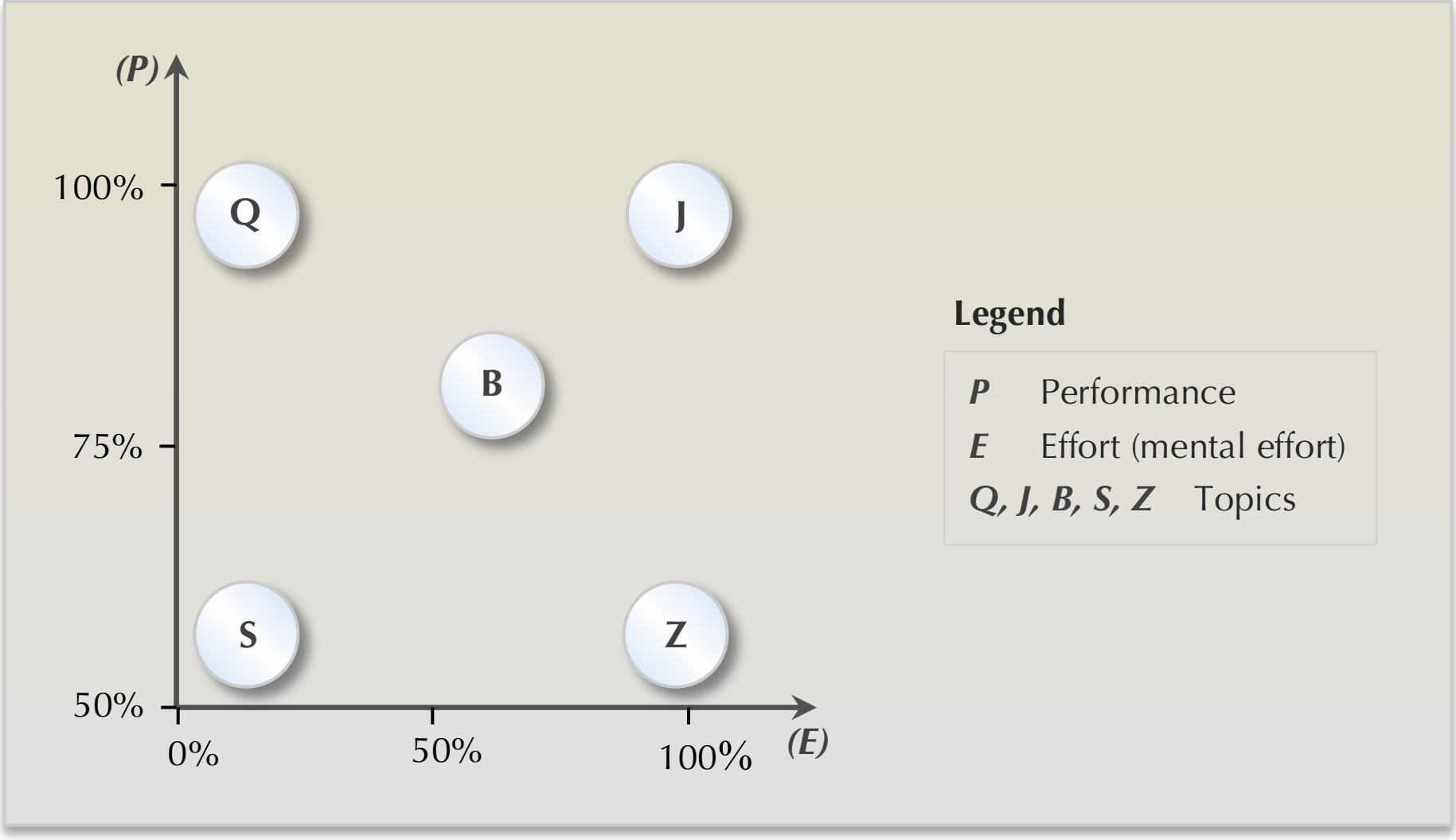

| Topic | Mental Effort | Performance | Cognitive Load Indications |
|---|---|---|---|
| Q | Low | High | Low Extraneous / (High Extraneous + Low Intrinsic) |
| J | High | High | High Germane |
| S | Low | Low | Low Germane |
| Z | High | Low | High Extraneous, Low Germane, High Intrinsic |
| B | Medium-High | Medium-High | Medium-High Germane |

**Figure 6**: Cognitive load profiling of topics based on learners' dual performance and mental effort scores (*P*, *E*) in a five-topic learning environment.

Specifically, it is observed that the majority of learners tend to achieve very well in Topics Q and J, although success in the latter comes after great effort. On the other hand, they perform poorly in Topic S without much effort, whereas in Topic Z performance is poor despite high effort; finally, they achieve above average in Topic B without putting in maximum effort.

Assuming that these patterns were retrieved from both trained maps, the instructor can proceed to determine a cognitive load profile for the learning environment. For example, high mental effort combined with high performance in Topic J indicates that germane cognitive load is most likely high. For Topic Q, low mental effort combined with high performance indicates either low extraneous cognitive load, or high extraneous cognitive load combined with low intrinsic cognitive load. Depending on these cognitive load indications, the instructor can make reliable assumptions regarding the learners' level of acquired knowledge and working memory load, and pinpoint instructional design weaknesses in the environment with high confidence.

The explanatory diagram in Figure 7 illustrates the organic relationship between the neurocomputational components of the learner model and the implications for cognitive load. Germane, extraneous and intrinsic cognitive load levels are indicated in relation to performance and mental effort scores in accordance with CLT. In particular, the germane ($G$) and extraneous ($E^*$) cognitive load axes are placed in the diagram relative to the performance ($P$) and mental effort ($E$) axes. An intrinsic cognitive load axis could not be depicted due to the complexity of its relationship to the other components of the diagram. Selected example points in this tightly interrelated diagram are placed on $G$ and $E^*$ axes to depict cognitive load levels for various combinations of performance and mental effort scores.

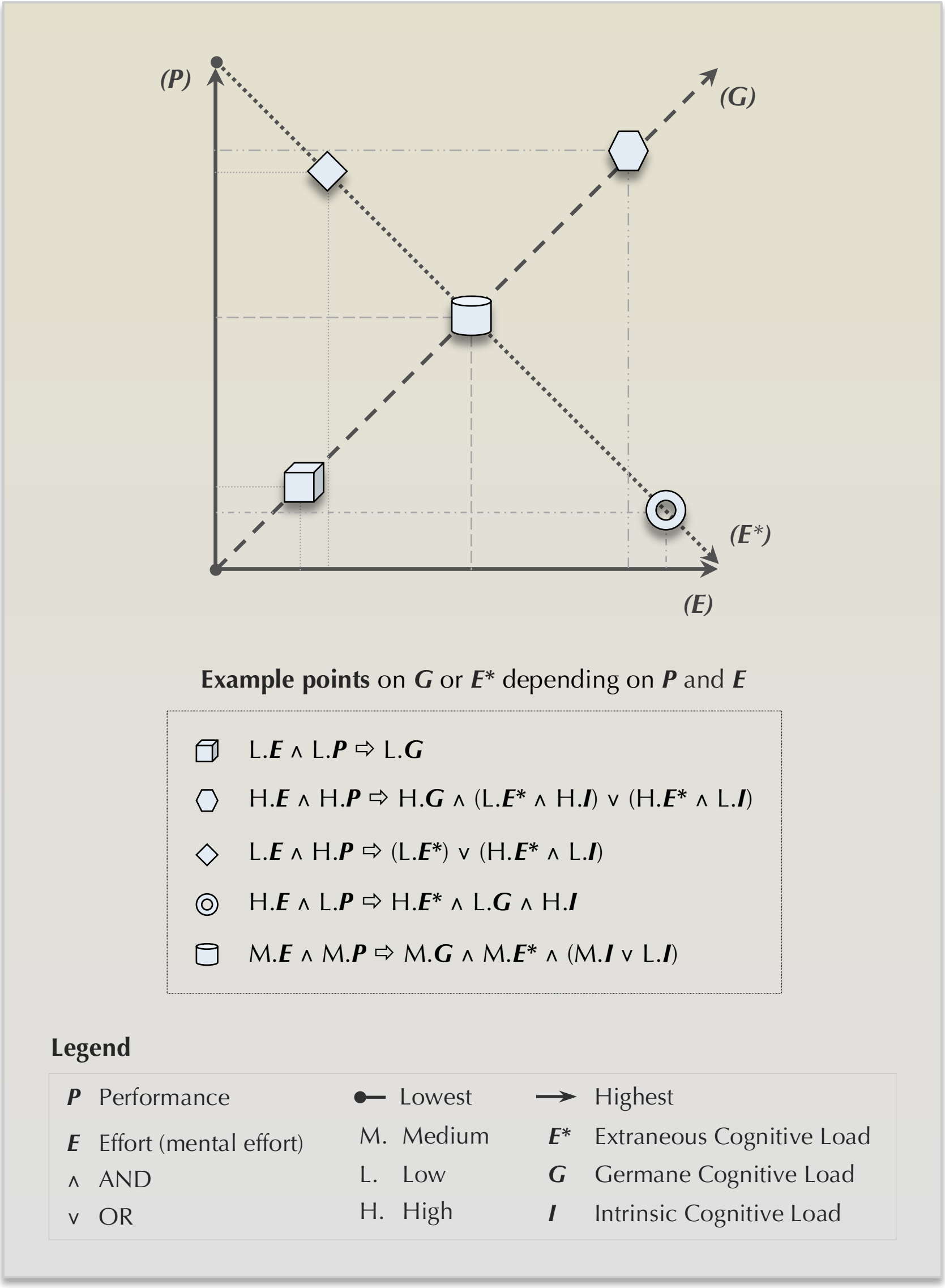

**Figure 7**: Learning environment collective cognitive load as a function of performance (*P*) and mental effort (*E*). Five interpretational indicators of cognitive load are shown as example points on two visible cross axes (*G* and *E**).

As depicted in Figure 7, low mental effort coupled with low performance indicates low germane cognitive load (the ‘cube’ example point). Similarly, high mental effort coupled with low performance indicates high extraneous, low germane and high intrinsic cognitive load (the ‘doughnut’ example point). In some cases, the graphical depiction is ambiguous due to the absence of the intrinsic cognitive load axis from the diagram; this signifies that the cognitive load indication is conditional on either the non-depicted intrinsic cognitive load level or the complexity of the relationship between types of cognitive load. For example, low mental effort coupled with high performance indicates either low extraneous cognitive load (the ‘rhomb’ example point), or high extraneous cognitive load combined with low intrinsic cognitive load (whose axis is not depicted in the diagram). Furthermore, high mental effort coupled with high performance indicates high germane cognitive load (the ‘hexagon’ example point) combined with either low extraneous load and high intrinsic load (axis not depicted) or high extraneous load and low intrinsic load (axis not depicted).

Crucially, the relationship between combined performance and mental effort scores, on the one hand, and specific levels of cognitive load, on the other, is independent of the validity of the learner model; this is due to the fact that the latter is a stand-alone model of domain-specific learning behaviour with no reference to specific interpretational assumptions from CLT. Therefore, even if the explanatory diagram (or framework) is substantially revised, the model can still be employed to provide updated indicators of cognitive load for both previously examined and new learning environments.

## 4 Implementation and Evaluation Aspects

### 4.1 SOM Vector Structure

The model consists of two artificial neural networks, PSOM and ESOM, which possess different semantic content (performance and mental effort, respectively) but an identical structure. Specifically, each neuron weight vector in either neural network: (a) encodes scores within the real-number interval [0,1], corresponding to topic performance or mental effort scores, and (b) contains a number of elements equal to the total number of topics in the learning environment (Figure 8).

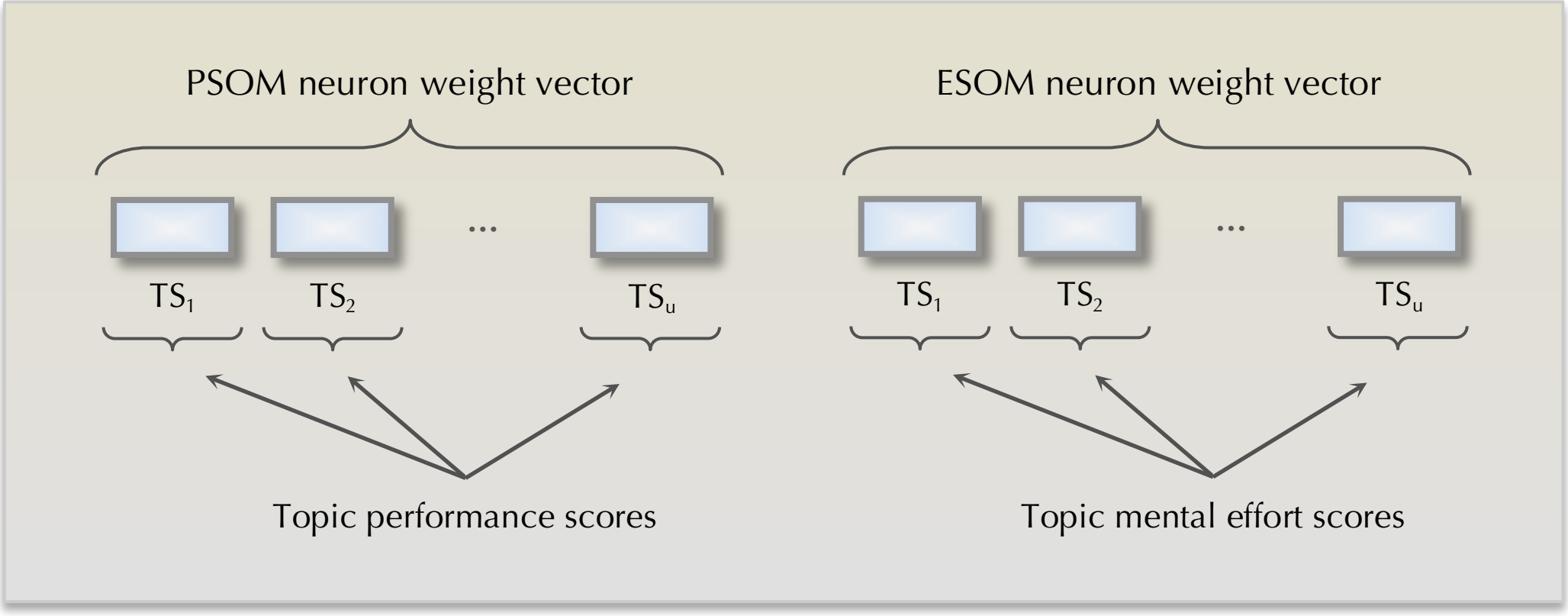

**Figure 8**: Neuron weight structure in PSOM and ESOM neural networks for a learning environment of *u* topics.

In terms of algorithmic analysis, if *m* is the cardinality (size) of the training set, *n* is the size of the network's neuron lattice, and *k* denotes the number of iterations of the self-organising and convergence part of the SOM formation algorithm, the complexity of the algorithm (for either PSOM or ESOM) at vector-level processing is generally $\Theta(m \cdot n \cdot k)$.

### 4.2 Validity of PSOM and ESOM

A real-life evaluation of the proposed dual SOM model would require sufficient training data gathered from prolonged experimentation. This would involve monitoring human learners and an instructor on a predesigned learning environment, running for a significant period of time until completion, as well as adequate class repetition, each time with a cohort of new students. Beyond the obvious practical challenges (e.g., the duration of the evaluation process) involved in this approach, there is also a significant methodological challenge regarding the model's evaluation against the behavioural trends of learners (i.e., their inclinations to follow certain sequences when completing topics in the given learning environment).

Specifically, the evaluation metric must measure success against the *behavioural trends* of a given learner group: the model must effectively reveal these trends, as doing so is instrumental to evaluating the instructional efficiency of the learning environment from a CLT perspective. This implies prior knowledge of the group's behavioural trends to serve as a reference point for the evaluation metric. In other words, the underlying behavioural trends must be known in order to measure how accurately the model depicts them in a trained SOM. In a real learning environment, it is particularly difficult to determine such behavioural inclinations with certainty; however, this is not the case in a simulated environment, where they can be imposed by design (and thus known prior to the evaluation process).

Accordingly, PSOM semantic encoding raises the question of whether the SOM neural network can feasibly detect and represent behavioural inclinations linked to hidden topic dependencies. Strong evidence indicates that this question can be answered in the affirmative based on the conclusions of an earlier investigation (Revithis, 2001), in which a SOM neural network with an identical structure to PSOM demonstrated the capacity to form statistically appropriate representations of both the training set (i.e., the domain behavioural space) and hidden topic dependencies within a five-topic learning environment. In that study, controlled computer SOM training simulations were performed using synthetic training data sets (i.e., artificial learners) containing predesigned embedded behavioural norms, reflected in distinct learner performance, followed by a systematic evaluation process. The SOM successfully depicted the predesigned behavioural norms of the learners. Revealing these norms is precisely what the PSOM component of the model is expected to accomplish, while the remaining part of the proposed model addresses the interpretation of performance (combined with mental effort depicted in ESOM) informed by CLT, as discussed in the preceding sections.

The implementation and evaluation of the ESOM neural network can be grounded entirely in the identically structured PSOM. Consequently, no new challenges or unforeseen technical issues are expected to arise when modelling mental effort using a representative training set drawn from the domain behavioural space. This is because ESOM possesses, by design, an identical neural network structure and semantic encoding format to PSOM: for each topic, PSOM retains performance scores while ESOM retains mental effort introspective scores. For the same reasons, the computational feasibility of ESOM is directly analogous to that of PSOM. A trained ESOM depicts a map of introspective mental effort patterns; due to the SOM topological ordering property, similar patterns are represented by neighbouring neuron nodes in the resulting two-dimensional ESOM map. Taken together, this structural and encoding symmetry between PSOM and ESOM substantiates both computational and representational validity across the two maps.

## 5 Current and Future Considerations

The novel model presented in this paper integrates a diverse set of cognitive and computational theories: a dual-SOM neural network is structured around an input-space encoding that enables the representation and analysis of learning behaviour in terms of working memory load within the framework of CLT. Such a synthesis entails several operational constraints and design considerations:

- Beyond the initial training of each SOM (PSOM, ESOM), the model requires continuous learning via SOM training sessions scheduled at regular intervals to remain responsive to behavioural and cognitive shifts in future cohorts. Applying a performance metric is both transparent to learners and well established, as topic assessment is a standard component of typical educational processes. On the other hand, supplying the model with introspective data is more challenging, as it requires learners to systematically complete self-evaluating tasks; from a learner's perspective, these tasks may appear unnecessary and relatively distant from the primary learning goal, potentially affecting the reliability of the metric. Consequently, future implementations should consider embedding introspective assessment as an integral part of the overall learning experience.
- As discussed in Section 3.2, the model does not rely on specific CLT assumptions to operate successfully: although designed to generate data that facilitates educational evaluations under CLT, it functions as a stand-alone learner model. However, translating the model's outputs into pedagogical action requires either human interpretation by an instructor or the development of an artificial agent designed for that purpose.
- The model is explicitly designed to provide the instructor of a learning environment with diagnostic insights regarding collective cognitive load and its implications for instructional design. This implies that a) there is no built-in mechanism to automatically recommend structural modifications to the learning environment, and b) any structural adjustments necessarily build upon the existing curriculum architecture; that is, the model cannot be meaningfully employed to redesign the instructional environment entirely from scratch without reference to its existing structure.

The following points address a number of important and anticipated challenges, and provide additional insights as well as further support in pursuing future work:

- Minimisation of selection bias in training data: By design, the training data set is restricted to successful trajectories from learners who clear an initial background-knowledge screening. This isolates true pedagogical friction within the curriculum from extraneous non-academic variables (e.g., personal or external dropout reasons) that could otherwise pollute the behavioural space.
- Bridging collective and individual cognitive load: The model captures both dimensions at distinct structural levels: individual learner vectors form the training set, while the resulting SOMs organise these into topological prototype clusters representing dominant behavioural patterns. Rather than averaging out individual differences, the competitive network maps distinct, statistically representative trajectories across the performance and mental effort spaces.

- Diagnostic scope and mapping specificity*:* The model is designed as a decision-support system that provides heuristic indicators to guide expert instructional design, rather than a closed-loop, fully automated diagnostic algorithm. Cases where identical performance and mental effort scores lead to divergent cognitive load interpretations are resolved by the instructor using neighbouring topological patterns, domain expertise, and external curricular context.
- Dependency graph refinement*:* Because all permitted trajectories adhere to the initial dependency graph and enrolment screening, systematic performance drops or effort spikes along specific valid orderings reveal implicit prerequisites between topics. The model isolates these hidden dependencies from individual ability through convergent statistical pattern clustering over repeated cohort completions.

Further considerations on the applicability of this model to educational psychology would eventually require real-life implementation projects in e-learning and cloud-oriented educational settings. Such implementations would entail a separate set of specifications and requirements as well as evaluation metrics.